\documentclass[%
 reprint, twocolumn,
 superscriptaddress,
 amsmath,amssymb,
 aps, prl,
floatfix,
]{revtex4-2}

\usepackage{graphicx}
\usepackage{dcolumn}
\usepackage{bm}
\usepackage{hyperref}
\hypersetup{colorlinks=true,citecolor=blue,filecolor=blue,urlcolor=blue,}
\usepackage[normalem]{ulem}

\newcommand{\lcdm}{$\Lambda$CDM}
\newcommand{\apjl}{ApJ}
\newcommand{\aj}{AJ}

\newcommand{\mnras}{MNRAS}

\newcommand{\aap}{A\&A}

\newcommand{\jcap}{JCAP}

\newcommand{\leftparbox}[2]{\parbox{#1}{\begin{flushleft} #2 \end{flushleft}}}

\newcommand{\threeonesig}[5][\pbwidth]{
\begin{equation}
\left.
 \begin{aligned}
#2 \\ #3 \\ #4
 \end{aligned}
\ \right\} \ \ \mbox{\text{\leftparbox{#1}{68\,\%,~#5}}}
\end{equation}
}

\newcommand{\dataplus}{\allowbreak+}
\newcommand{\mksym}[1]{\ifmmode {\rm #1}\else #1\fi}
\newcommand{\planck}{{Planck}}
\newcommand{\lya}{Ly$\alpha$}
\newcommand{\kmsMpc}{km\,s$^{-1}$Mpc$^{-1}$}

\begin{document}

\title{Testing Low-Redshift Cosmic Acceleration with Large-Scale Structure}

\author{Seshadri Nadathur}
 \email{seshadri.nadathur@port.ac.uk}
 \affiliation{Institute of Cosmology and Gravitation, University of Portsmouth, Burnaby Road, Portsmouth, PO1 3FX, United Kingdom}
\author{Will J. Percival}%
 \affiliation{Waterloo Centre for Astrophysics, Department of Physics and Astronomy, University of Waterloo, 200 University Avenue West, Waterloo, Ontario N2L 3G1, Canada}
 \affiliation{Perimeter Institute for Theoretical Physics, 31 Caroline Street North, Waterloo, Ontario N2L 2Y5, Canada}
\author{Florian Beutler}
 \affiliation{Institute of Cosmology and Gravitation, University of Portsmouth, Burnaby Road, Portsmouth, PO1 3FX, UK}
\author{Hans A. Winther}%
 \affiliation{Institute of Theoretical Astrophysics (ITA), University of Oslo, P.O. Box 1029, Blindern 0315, Oslo, Norway}

\date{\today}

\begin{abstract}
We examine the cosmological implications of measurements of the void-galaxy cross-correlation at redshift $z=0.57$ combined with baryon acoustic oscillation (BAO) data at $0.1<z<2.4$. 
We find direct evidence of the late-time acceleration due to dark energy at $>10\sigma$ significance from these data alone, independent of the cosmic microwave background and supernovae. Using a nucleosynthesis prior on $\Omega_bh^2$, we measure the Hubble constant to be $H_0=72.3\pm1.9$ \kmsMpc{} from BAO+voids at $z<2$, and $H_0=69.0\pm1.2$ \kmsMpc{} when adding Lyman-$\alpha$ BAO at $z=2.34$, both independent of the CMB. Adding voids to CMB, BAO and supernova data greatly improves measurement of the dark energy equation of state, increasing the figure of merit by $>40\%$, but remaining consistent with flat $\Lambda$ cold dark matter.
\end{abstract}

\maketitle

\emph{Introduction}.---The standard \lcdm{} model of cosmology requires a negative pressure dark energy (DE) component responsible for the observed late-time acceleration of the expansion rate that is theoretically still not well understood. In \lcdm, the tension between the values of the Hubble constant $H_0$ obtained from the local distance ladder \cite{Riess:2019} and lensing time-delay \cite{H0LiCOW:2019H0} methods, and from the cosmic microwave background (CMB) by Planck \cite{Planck:2018params}, is 4-6$\sigma$ \cite{Verde:2019}. These are among the biggest challenges to our model of the Universe.

Measurement of the expansion history of the Universe at low redshifts provides observational tests key to both issues. Using type Ia supernovae (SNe) as standard candles originally established the acceleration due to DE \cite{Perlmutter:1999,Riess:1998}. However, the large-scale structure (LSS) of the Universe provides an independent competitive test of the expansion rate, through observation of baryon acoustic oscillations (BAO) in galaxy surveys at different redshifts \cite{Addison:2013}. Recently, Reference \cite{Nadathur:2019c} applied the Alcock-Paczynski (AP) \cite{Alcock:1979} test to a new measurement of the void-galaxy cross-correlation, and showed how the combination of this observable with BAO sharpens the distance scale and expansion rate measurements achievable from existing LSS surveys. These measurements may be calibrated relative to the sound horizon scale in the early Universe, determined either from the CMB or independently using big bang nucleosynthesis (BBN) and the primordial deuterium abundance \cite{Addison:2018}, or used without any external calibration, providing direct independent tests of cosmic acceleration.

In this Letter, we examine the constraints on the low-redshift expansion history provided by the latest BAO data, combined for the first time with the new void-galaxy cross-correlation results. We first show that within flat \lcdm, voids increase the value of $H_0$ obtained independent of the CMB. Combined with BAO data at $z<2$, this favours the local distance ladder value of $H_0$, though Lyman-$\alpha$ (\lya) BAO at higher $z$ pull to lower $\Omega_m$ and $H_0$, making the result compatible with Planck. In more general spatially curved models BAO and voids together provide direct evidence of late-time acceleration at a much higher significance than SNe. Finally, we combine all datasets to obtain the best current measurements of the DE equation of state and the tightest observational constraints on DE models.

\begin{figure*}
\includegraphics[width=0.95\linewidth]{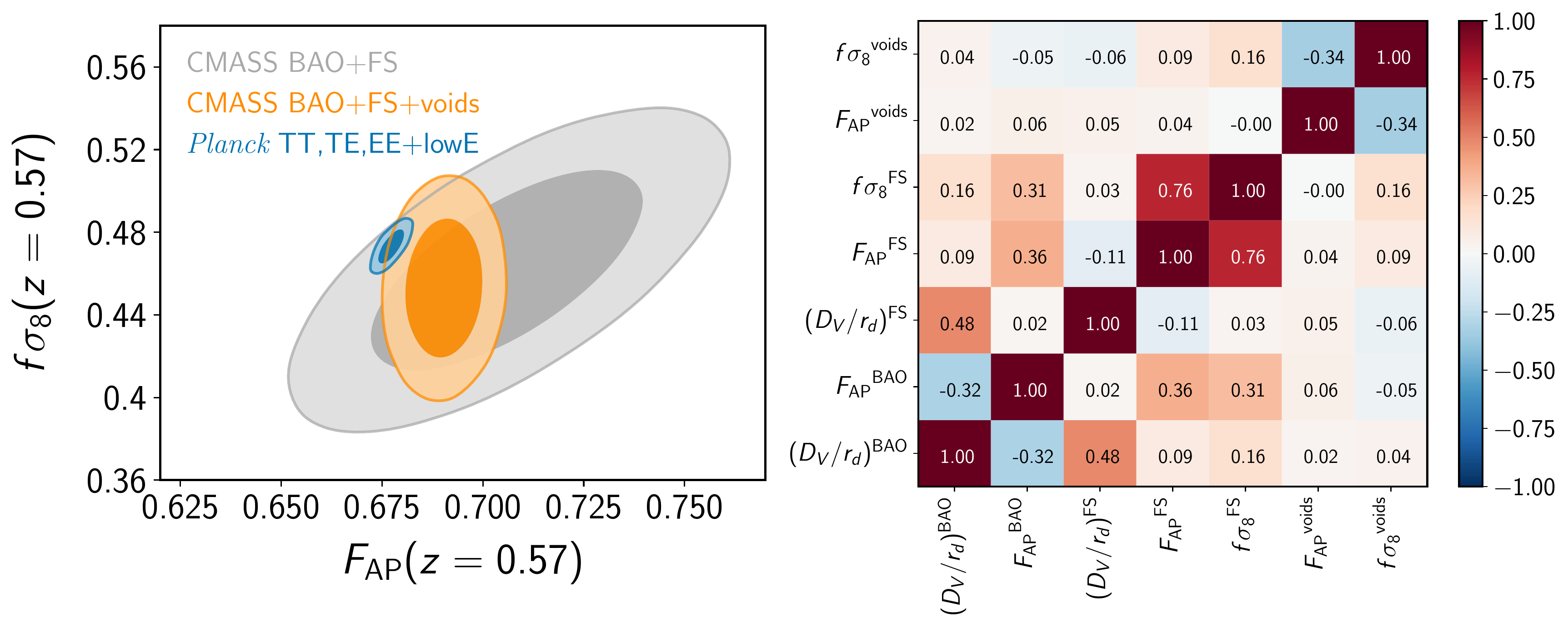}
\caption{\emph{Left}: Measurement of the Alcock-Paczynski parameter $F_\mathrm{AP}=D_MH/c$ and growth rate $f\sigma_8$ from the BOSS CMASS sample (68\% and 95\% contours) \cite{Gil-Marin:2016a, Gil-Marin:2016b, Nadathur:2019c}. Planck constraints are extrapolations from CMB fits assuming \lcdm. \emph{Right}: Correlation coefficients for the constituent BAO, FS and void measurements in this sample, estimated from mocks \cite{Nadathur:2019c}.} 
\label{fig:F_AP and growth}
\end{figure*}

\emph{Methods and data}.---BAO analyses measure the scale of the sharp feature in the correlation function (or oscillations in the power spectrum) of galaxies, quasars or the \lya{} forest. The BAO scale is set by the sound horizon $r_d$ at the drag epoch $z_d$ when photons and baryons decouple, $r_d = \int^\infty_{z_d}c_s(z)/H(z)\mathrm{d}z$, where $c_s(z) = 3^{-1/2}c\left[1+\frac{3}{4}\rho_b(z)/\rho_\gamma(z)\right]^{-1/2}$ is the sound speed in the photon-baryon fluid. In practice, the observed tracer redshifts and angles on the sky must be converted to distances by adopting a fiducial cosmological model, and the analysis measures the ratio of the observed BAO scale to that predicted in the fiducial model. An angle-averaged isotropic fit therefore measures $D_V(z)/r_d$ \cite{Eisenstein:2005, Padmanabhan:2008, Ross:2017, Beutler:2017}, where
\begin{equation}
    \label{eq:D_V}
    D_V(z) = \left[czD_M^2(z)/H(z)\right]^{1/3},
\end{equation}
$D_M(z)$ is the transverse comoving distance \cite{Hogg:1999}, and $H(z)=H_0\left[\Omega_m(1+z)^3+\Omega_K(1+z)^2+\Omega_\Lambda\right]^{1/2}$ with $\Omega_m$, $\Omega_K$ and $\Omega_\Lambda$ the energy densities of matter, curvature, and DE, respectively (neglecting the energy density in radiation). When BAO features along and perpendicular to the line-of-sight direction can be individually resolved by an anisotropic analysis, they measure $H(z)r_d$ and $D_M(z)/r_d$ respectively. The AP test then requires the same size of the BAO feature along and across the line of sight, and constrains the combination $F_\mathrm{AP}(z)\equiv D_M(z)H(z)/c$. 

The void-galaxy cross-correlation function $\xi_{\rm vg}$ provides a complementary test via the distribution of galaxies around the centres of low-density void regions \cite{Lavaux:2012, Hamaus:2015, Hamaus:2016, Nadathur:2019c}. As in the case of the galaxy autocorrelation, anisotropy in $\xi_{\rm vg}$ is sourced by both redshift-space distortions (RSDs) due to galaxy peculiar velocities, and due to the AP effect: i.e., differences between the assumed fiducial model and the true cosmology, which cause the ratio $\epsilon=F_\mathrm{AP}^\mathrm{true}/F_\mathrm{AP}^\mathrm{fid}$ to differ from 1. In the autocorrelation, these two effects are degenerate and hard to separate \cite{Ballinger:1996, Matsubara:1996}, a difficulty compounded by the fact that non-linearities in the coupling of density and velocity fields limit the range of scales over which RSD models can be applied, e.g. \cite{Scoccimarro:2004, Matsubara:2008, Taruya:2010, Reid:2011, Jennings:2011a, Ivanov:2019a, D'Amico:2019, Colas:2019}. In contrast, RSD contributions to $\xi_{\rm vg}$ can be accurately modelled by linear perturbation theory down to very small scales \cite{Nadathur:2019a} (after correcting for systematic biases in void selection in redshift space using a reconstruction technique closely related to that used for BAO \cite{Nadathur:2019b}), and the RSD and AP terms produce distinctive and easily separable signatures in the quadrupole moment of $\xi_{\rm vg}$, at scales of $\sim20$-$30\;h^{-1}$Mpc \cite{Nadathur:2019c}. As a result, the measured anisotropy of the void-galaxy cross-correlation provides a $\sim1\%$ measurement of $F_\mathrm{AP}$, exceeding the precision that can be obtained from BAO by a factor of $\sim4$ \cite{Nadathur:2019c}. Reference~\cite{Nadathur:2019c} used tests on mocks to demonstrate that systematic errors in this measurement are negligible at this level. This constraint arises from the AP test applied to the shape of special objects (i.e., voids) in the Universe, and therefore represents a gain in information over any analyses that do not isolate voids. This precision in $F_\mathrm{AP}$ means that the combination of BAO and void-galaxy cross-correlation breaks the degeneracy between $H(z)r_d$ and $D_M(z)/r_d$ and significantly reduces the uncertainties in each.

We use anisotropic BAO measurements of $D_M(z)/r_d$ and $H(z)r_d$ from the Baryon Oscillation Spectroscopic Survey (BOSS; \cite{Dawson:2013}) final DR12 data release \cite{Alam:2017}, and from the eBOSS DR14 \lya{} BAO measurement \cite{Blomqvist:2019,deSainteAgathe:2019}. For BOSS we use the ``consensus" results, which include both the post-reconstruction BAO-only fits and the full-shape (FS) analyses in three overlapping redshift bins, $z=0.38,\,0.51,\,0.61$. For eBOSS \lya{} we use the combined \lya{} autocorrelation and \lya$\times$quasar results, at effective redshift $z=2.34$. To these we add the isotropic fits to $D_V(z)/r_d$ from the 6dF Galaxy Survey (6dFGS; \cite{Beutler:2011hx}), the SDSS Main Galaxy Sample (MGS; \cite{Ross:2015}), and the eBOSS DR14 quasar sample \cite{Ata:2018}, at redshifts $0.106,\,0.15$ and $1.52$ respectively. 

\begin{figure*}
\includegraphics[width=0.95\linewidth]{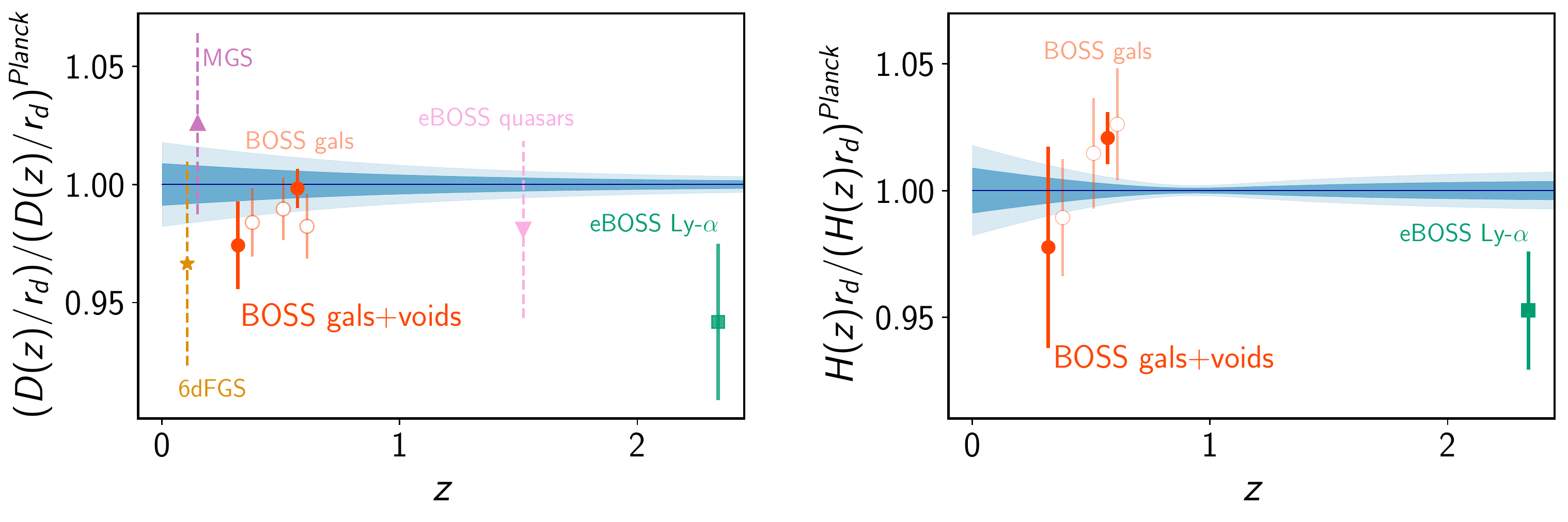}
\caption{Distance scale $D(z)/r_d$ (left) and Hubble rate $H(z)r_d$ (right) measurements from different surveys, relative to their values in the best-fit $\Lambda$CDM model from \planck{}. In the left panel, solid error bars (BOSS and eBOSS \lya) denote $D_M/r_d$. Points with dashed error bars (6dFGS, MGS and eBOSS quasars) measure only $D_V/r_d$. The orange open circles are the consensus BAO+FS results from BOSS DR12 galaxies only. The red filled circles are from the same data, but rebinned and including void information at $z=0.57$. Uncertainties are correlated for points in common between both panels. Shaded regions show \planck{} uncertainties.}
\label{fig:distances}
\end{figure*}

In addition to this, we use the void-galaxy cross-correlation results from Nadathur {\it et al.}~\cite{Nadathur:2019c}, who presented joint constraints on $D_M(z)/r_d$ and $H(z)r_d$ from the void-galaxy measurement and its combination with BAO \cite{Gil-Marin:2016a, Cuesta:2016} and FS \cite{Gil-Marin:2016b} results from BOSS DR12, including the cross-covariance between methods, shown in Fig.~\ref{fig:F_AP and growth}. References \cite{Nadathur:2019c, Gil-Marin:2016a, Cuesta:2016, Gil-Marin:2016b} present results for BOSS DR12 in two independent bins at effective redshifts $z=0.32$ and $z=0.57$, corresponding to the LOWZ and CMASS subsamples respectively. The void-galaxy measurement has only been made for the CMASS subsample, so the gains in precision only apply to that redshift bin.

The results for $D_{(M/V)}/r_d$ and $Hr_d$ are summarized in Fig.~\ref{fig:distances}. In what follows, when combining the 6dFGS, MGS, eBOSS quasar and eBOSS \lya{} data points with the consensus BOSS results in three redshift bins, we refer to the full dataset collectively as ``BAO". For the ``BAO+voids" dataset we replace the BOSS consensus results above with those from Reference~\cite{Nadathur:2019c} in two redshift bins instead. We use the identifier ``voids" to refer to the constraint $F_\mathrm{AP}=0.6859\pm 0.0071$ obtained from the void-galaxy measurement alone, without BAO.

Finally, for some analyses we use two additional external datasets: the Pantheon sample of type Ia SNe \cite{Scolnic:2018}, and the CMB temperature and polarization power spectra from Planck 2018 \cite{Planck:2018params}. We explore the parameter spaces of models using Markov chain Monte Carlo based on the \texttt{CosmoMC} code \cite{Lewis:2002} and examine chains using \texttt{GetDist} \cite{Lewis:GetDist}.

\emph{Hubble constant}.---We start with the most restrictive class of flat \lcdm{} models with cosmological constant DE and standard recombination physics. $D_M(z)/r_d$ and $H(z)r_d$ results from BAO and voids then provide constraints in the two-parameter $\left(\Omega_m, H_0r_d\right)$ plane, with voids alone providing a constraint $\Omega_m=0.35\pm0.03$. Measurement of $D/H$ \cite{Cooke:2018} combined with BBN theory can be converted to a prior on the baryon density $\Omega_bh^2$ independent of CMB anisotropy information. We adopt the conservative prior $\Omega_bh^2=0.0222\pm0.0005$, motivated by \cite{Cooke:2018} but including an increased uncertainty and slightly shifted mean value to account for the systematic differences in $\Omega_bh^2$ values obtained using theoretical and empirical estimates of the key $d(p,\gamma)^3\mathrm{He}$ reaction rate in BBN \cite{Marcucci:2016,Cooke:2018,Cuceu:2019}. This is sufficient to break the $r_d-H_0$ degeneracy and determine $H_0$ independent of the CMB \cite{Addison:2018}.

\begin{figure}
\includegraphics[width=0.9\linewidth]{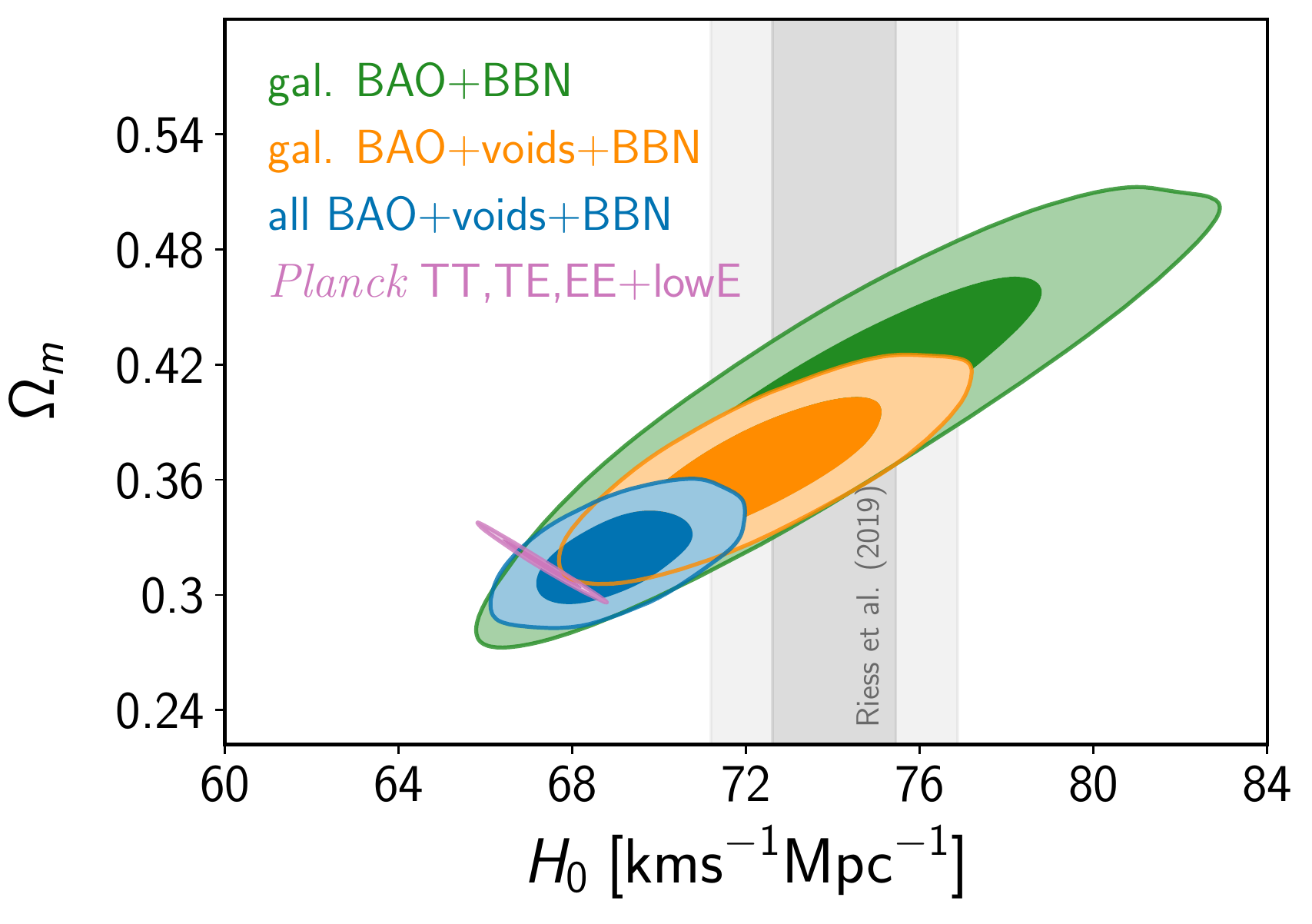}
\caption{Marginalized 68.3\% and 95.2\% contours for $\Omega_m$ and $H_0$ in \lcdm{} from Planck, and from BAO and voids combined with a BBN prior on $\Omega_bh^2$, independent of the CMB. `Gal. BAO' refers to all BAO measurements at $z<2$, favouring high $H_0$. 'All BAO' includes the \lya{} result at $z=2.34$, which pulls the final value down to $H_0=69.0\pm1.2$ \kmsMpc. Grey bands show the local distance ladder result of Reference \cite{Riess:2019}.}
\label{fig:H0}
\end{figure}

Reference \cite{Addison:2018} noted a $2.4\sigma$ tension between BAO results from galaxies and quasars at $z<2$ (which we refer to as `galaxy BAO') and \lya{} BAO at $z=2.34$. Although this has reduced to $1.9\sigma$ in the DR14 \lya{} results \cite{Cuceu:2019}, we first examine constraints from the two sets separately. As shown in Fig.~\ref{fig:H0}, galaxy BAO alone favour a high Hubble rate, albeit with large uncertainties: $H_0=73.7^{+3.0}_{-3.9}\;\mathrm{kms}^{-1}$Mpc$^{-1}$ and with a strong degeneracy between $\Omega_m$ and $H_0$. The void $F_\mathrm{AP}(z=0.57)$ measurement greatly reduces this uncertainty, giving
\begin{equation}
    \label{eq:H0galBAO+voids}
    H_0 = 72.3\pm 1.9\;\mathrm{km\,s}^{-1}\mathrm{Mpc}^{-1},\;\;(\mathrm{gal.~BAO+voids+BBN})
\end{equation}
a 2.6\% measurement independent of the CMB, consistent with the local distance ladder value $H_0=74.03\pm1.42$~\kmsMpc \cite{Riess:2019} but in $\sim2.5\sigma$ tension with the Planck result $H_0=67.36\pm0.54$~\kmsMpc \cite{Planck:2018params}. However, the low \lya{} measurement of $H(z)r_d$ favours low $\Omega_m$ (see \cite{Cuceu:2019}), so adding this constraint pulls the central $H_0$ value low again. (The same is true if \lya{} is replaced by other datasets such as Dark Energy Survey Year 1 clustering and weak lensing \cite{DES-H0:2018} which also favour low $\Omega_m$.) Thus, the final joint constraint we obtain is
\begin{equation}
    \label{eq:H0BAO+voids}
    H_0 = 69.0\pm 1.2\;\mathrm{km\,s}^{-1}\mathrm{Mpc}^{-1},\;\;(\mathrm{all~BAO+voids+BBN}).
\end{equation}
This represents a $1.7\%$ `early-type' measurement of $H_0$ from LSS, independent of the CMB. The final joint value is consistent with Planck and in $\sim2.7\sigma$ tension with the local distance ladder. Compared to result of \cite{Cuceu:2019}, $H_0=67.6\pm1.1$ \kmsMpc, the addition of the voids $F_\mathrm{AP}$ constraint has shifted the $H_0$ value up by $\sim1\sigma$ by pulling towards larger $\Omega_m$, counteracting the effect of \lya. Our value is similarly $\sim1.2\sigma$ higher than that of \cite{DES-H0:2018}. However, it should be noted that the low $\Omega_m$ value from \lya{} remains in some degree of tension with that from galaxy BAO + voids.

\begin{figure}
\includegraphics[width=0.9\linewidth]{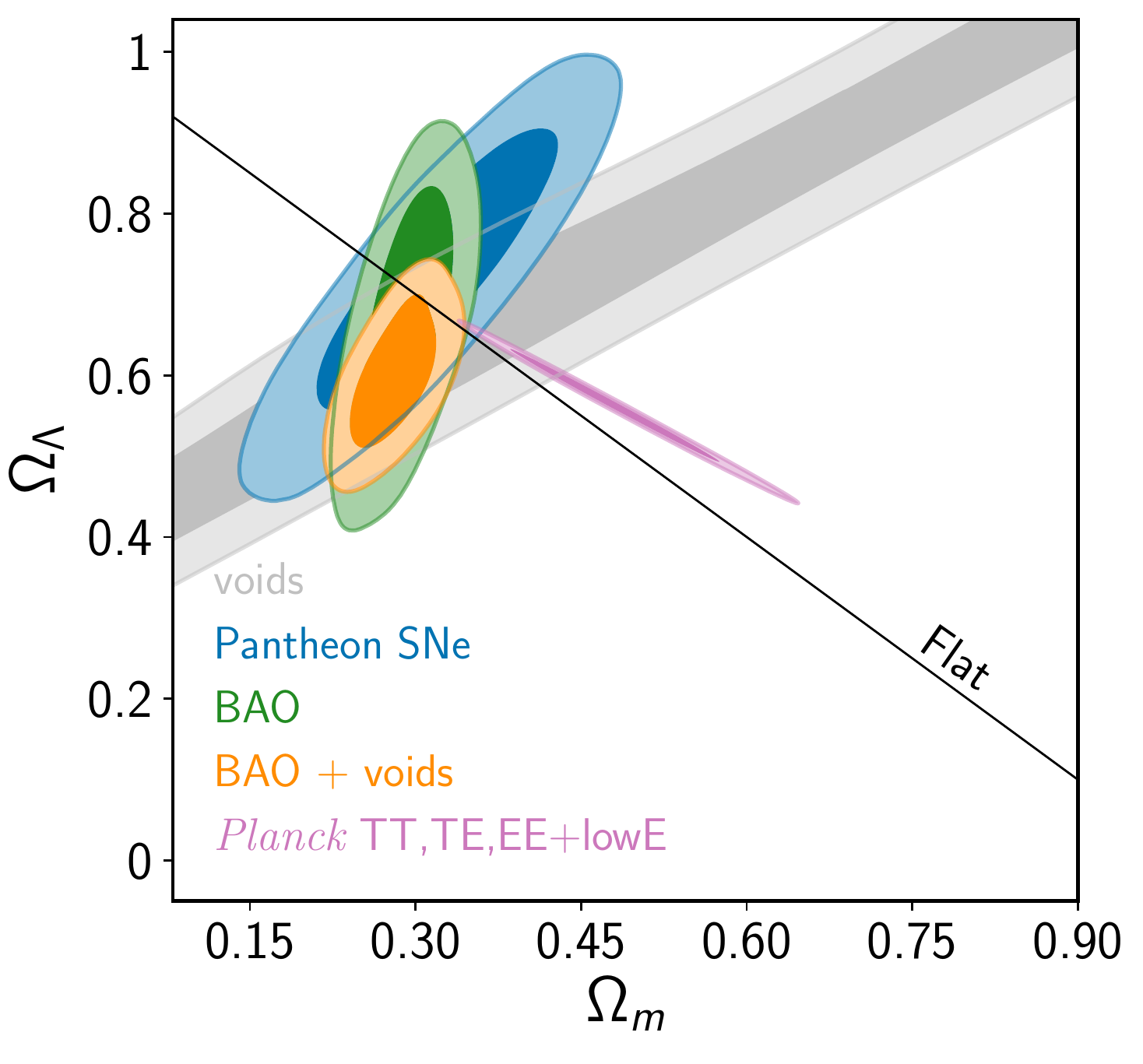}
\caption{Marginalised constraints on $\Omega_m$ and $\Omega_\Lambda$ obtained from BOSS voids, Pantheon SNe, BAO and Planck CMB, assuming $w=-1$. The line indicates spatially flat models. BAO+voids give $\Omega_\Lambda=0.600\pm 0.058$, a $>10\sigma$ detection of acceleration.}
\label{fig:concordance}
\end{figure}

\emph{Late-time acceleration}.---Relaxing the restrictive assumption of flatness, we examine the direct evidence for late-time acceleration due to DE. We assume an FLRW background with fixed $w=-1$ but the values of $\Omega_\Lambda$ and $\Omega_K$ left free. Then BAO measurements constrain the three parameters $\left(\Omega_m, \Omega_\Lambda, r_dH_0\right)$, while voids alone constrain $\left(\Omega_m, \Omega_\Lambda\right)$. As the sound horizon only appears in the degenerate combination $r_dH_0$, we need not calculate $r_d$ and require no knowledge of the physics of the early Universe beyond the fact that $\Omega_b>0$, so that a baryonic oscillation feature exists and sets a common scale measured by the various BAO experiments at different redshifts \cite{Addison:2013}.

The resulting marginalized limits on $\Omega_m$ and $\Omega_\Lambda$ are shown in Fig.~\ref{fig:concordance} compared to Planck and SNe constraints. BAO alone require $\Omega_\Lambda=0.675^{+0.11}_{-0.088}$, primarily driven by the combination of the BOSS at $z\sim0.5$ and eBOSS \lya{} at $z=2.34$ effectively breaking the degeneracy between $\Omega_m$ and $r_dH_0$. The single void data point of $F_\mathrm{AP}(z=0.57)$ also restricts viable models to a narrow band in the $\left(\Omega_m,\Omega_\Lambda\right)$ plane. With the weak prior $\Omega_m\ge0$, voids \emph{alone} require $\Omega_\Lambda>0$ at over $99.99\%$ confidence.

The combination of BAO with voids gives
\begin{equation}
    \label{eq:LambdaBAO+voids}
    \Omega_\Lambda = 0.60\pm 0.058,\;\;(\mathrm{BAO+voids})
\end{equation}
providing direct geometrical evidence of late-time acceleration due to DE at well over $10\sigma$ statistical significance. This far exceeds the precision of, but agrees with, the Pantheon SNe value $\Omega_\Lambda=0.73^{+0.12}_{-0.11}$. This result is based only on the assumption of statistical isotropy and comparison of the apparent size of the BAO standard ruler at different redshifts, so is completely independent of the CMB. 

As shown in Fig.~\ref{fig:concordance}, Planck temperature and polarization data alone favour a closed Universe with $\Omega_m\simeq0.5$ and $\Omega_K\simeq-0.04$ \cite{Planck:2018params,Handley:2019, DiValentino:2019} (although the significance of $\Omega_K<0$ depends on the CMB likelihood \cite{Efstathiou:2019} and is reduced by Planck lensing \cite{Planck:2018lensing}). Both BAO+voids and SNe independently disfavour this closed model. 

\begin{figure*}
\includegraphics[width=0.44\linewidth]{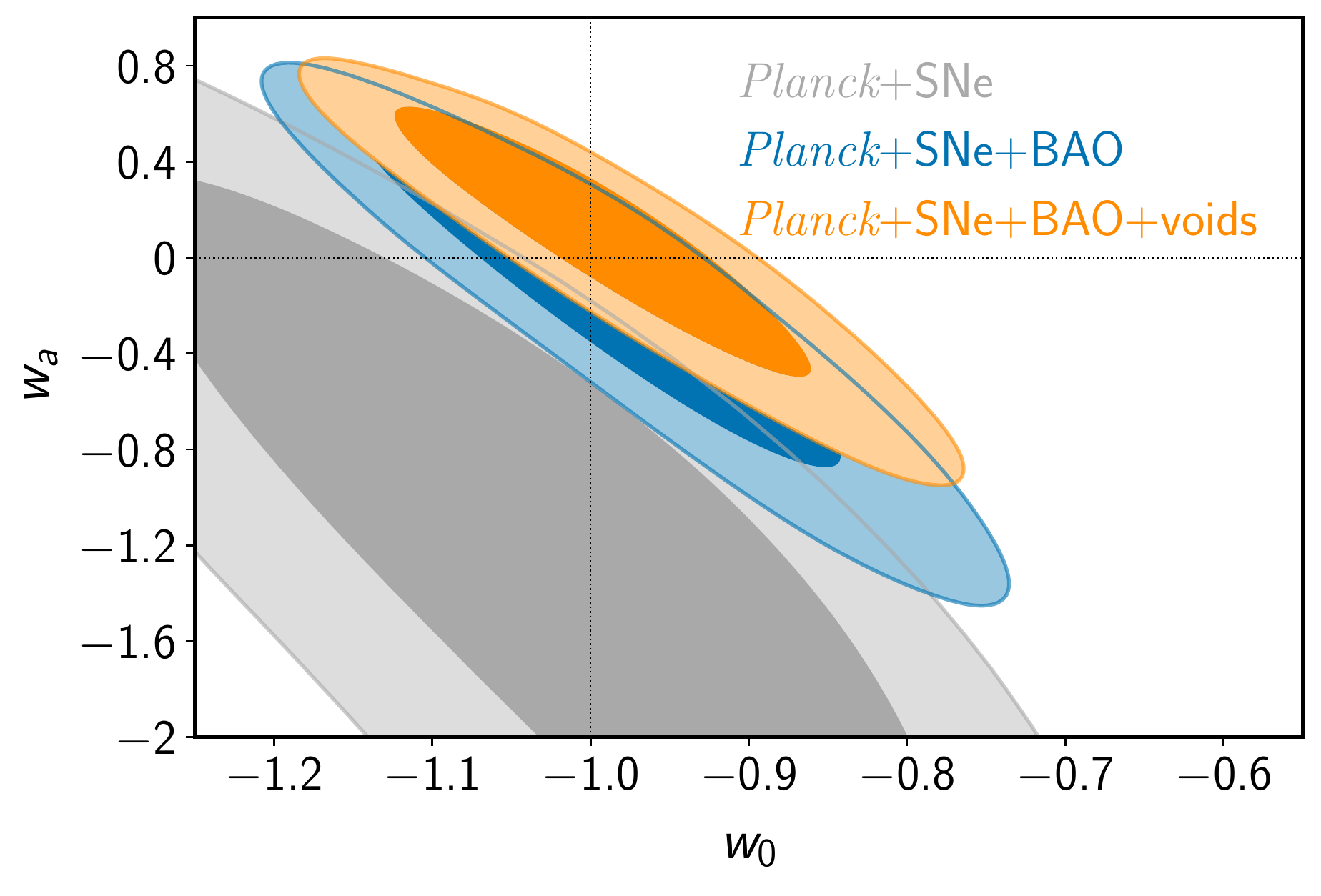}
\includegraphics[width=0.45\linewidth]{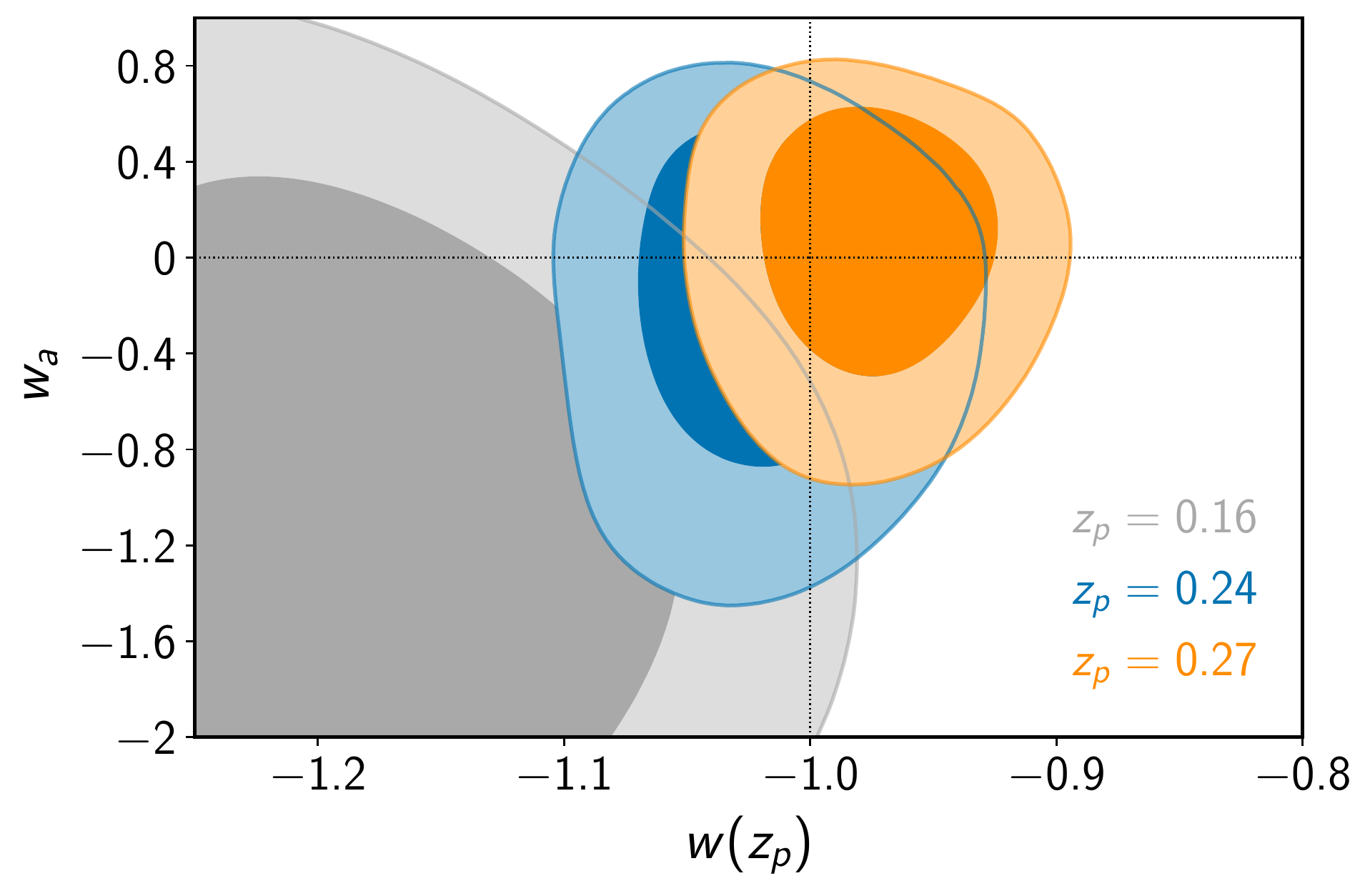}
\caption{\emph{Left}: Constraints on the DE equation of state parameters $w_0=w(z=0)$ and $w_a$ in non-spatially flat models. \emph{Right}: The same constraints represented in terms of the equation of state $w(z_p)$ at the pivot redshift, $z_p$, and $w_a$. The void $F_\mathrm{AP}$ measurement leads to a 43\% increase in the figure of merit over the best combination of other data.}
\label{fig:EOS}
\end{figure*}

\emph{DE equation of state}.---We now consider models in which, in addition to the base 6 parameters of \lcdm, $w$ is allowed to vary with redshift according to the popular $\left(w_0, w_a\right)$ parametrization \cite{Chevallier:2001, Linder:2003} and $\Omega_K$ is left free, denoted $ow_0w_a$CDM. Low-redshift measurements of $D_M$ and $H$, and particularly the combination $F_\mathrm{AP}$, are crucial to breaking degeneracies in fits of this model to CMB and SNe data; therefore, the addition of BAO and void-galaxy results gives a great deal of information. 

We compare constraints on the $ow_0w_a$CDM model obtained from fitting the Planck+SNe, Planck+SNe+BAO and Planck+SNe+BAO+voids data combinations. In both the latter two cases, BAO from all surveys are included, although in practice the effect of BOSS dominates due to its much higher precision. For BOSS, we also include the measured constraints on $f(z)\sigma_8(z)$, where $f=\mathrm{d}\ln\delta/\mathrm{d}\ln a$ is the linear growth rate and $\sigma_8(z)=\sigma_8\left[\delta(z)/\delta(0)\right]$ with $\sigma_8$ the rms linear mass fluctuations in $8\;h^{-1}$Mpc spheres at $z=0$. The model has nine free cosmological parameters, for which priors are taken as in Reference \cite{Planck:2018params}. Results are shown in Fig.~\ref{fig:EOS}. The additional void $F_\mathrm{AP}$ constraint significantly reduces the errors, resulting in 
\threeonesig{w_0=-0.984^{+0.076}_{-0.097},}{w_a=0.05^{+0.44}_{-0.29},}{\Omega_K=0.0033^{+0.0034}_{-0.0041},}
{\mksym{$ow_0w_a$CDM:\;Planck}\dataplus\mksym{SNe}\dataplus\mksym{BAO}\dataplus\mksym{voids} \label{eq:owwaCDM}}
for the best combination. 

To quantify the information gain from individual datasets, we adopt the Dark Energy Task Force definition of the figure of merit \cite{Albrecht:2006}, 
\begin{equation}
    \label{eq:FoM}
    \mathrm{FoM}=\left[\sigma(w_p)\sigma(w_a)\right]^{-1}\,
\end{equation} 
where $\sigma(w_p)$ is the uncertainty in $w(z_p)=w_0+\frac{z_p}{1+z_p}w_a$ at the pivot redshift $z_p$ at which this error is minimised, given by $(1+z_p)^{-1}=1+\langle\delta w_0\delta w_a\rangle/\sigma^2_{w_a}$. For the Planck+SNe, Planck+SNe+BAO and Planck+SNe+BAO+voids cases the FoM values obtained are 10.9, 58.1 and 82.9 respectively. Thus the addition of the single void $F_\mathrm{AP}$ measurement represents a 43\% improvement in the FoM, achieved without the requirement of any new observational data. The pivot redshift for this case is $z_p=0.27$, and $w(z_p)=-0.974\pm 0.032$. Additionally requiring spatial flatness $\Omega_K=0$ improves constraints in all cases: we obtain $\mathrm{FoM}=39.1$, $108.1$ and $137.0$ for Planck+SNe, Planck+SNe+BAO and Planck+SNe+BAO+voids respectively. For the best Planck+SNe+BAO+voids case we find $w_0=-0.937\pm 0.074$ and $w_a=-0.22^{+0.28}_{-0.25}$, corresponding to $w(z_p)=-0.994\pm 0.027$ at pivot redshift $z_p=0.35$. 

\emph{Conclusions.}---Our results are consistent with the standard \lcdm{} model of a spatially flat Universe with a cosmological constant $\Lambda$. They represent the tightest constraints on deviations from this model, and the best measurement of dark energy, from any current data. They highlight the power of LSS data as a precise probe of the late-time acceleration, exceeding that of SNe. They also highlight, for the first time, the large gain in information provided by the inclusion of void-galaxy cross-correlation results, and the synergy between these and BAO. Void measurements are enabled by the same galaxy survey data as BAO, but represent information that cannot otherwise be obtained from the galaxy power spectrum or higher moments. The gains shown here are thus also more generally applicable to other cosmological models, such as those with non-minimal neutrino masses. A full quantification of the achievable information gain, especially from near-future surveys DESI \cite{DESI:2016} and Euclid \footnote{\url{https://www.euclid-ec.org/}}, is of immediate importance for future work.

\vspace{1em}S.N. was supported by UK Space Agency grant ST/N00180X/1. F.B. received support from a Royal Society University Research Fellowship. Computational work was performed on the UK {\small SCIAMA} High Performance Computing cluster supported by the ICG, SEPNet and the University of Portsmouth. This work made use of public catalogues from the SDSS-III and SDSS-IV. Funding for SDSS has been provided by the Alfred P. Sloan Foundation, the Participating Institutions, the National Science Foundation, and the U.S. Department of Energy Office of Science. The SDSS website is \url{http://www.sdss.org/}.

SDSS-IV is managed by the Astrophysical Research Consortium for the Participating Institutions of the SDSS Collaboration including the University of Arizona, the Brazilian Participation Group, Brookhaven National Laboratory, Carnegie Mellon University, University of Florida, the French Participation Group, the German Participation Group, Harvard University, the Instituto de Astrofisica de Canarias, the Michigan State/Notre Dame/JINA Participation Group, Johns Hopkins University, Lawrence Berkeley National Laboratory, Max Planck Institute for Astrophysics, Max Planck Institute for Extraterrestrial Physics, New Mexico State University, New York University, Ohio State University, Pennsylvania State University, University of Portsmouth, Princeton University, the Spanish Participation Group, University of Tokyo, University of Utah, Vanderbilt University, University of Virginia, University of Washington, and Yale University.

\bibliography{refs}
\bibliographystyle{mod-apsrev4-2.bst}

\end{document}